# On the origin of molecular oxygen on the surface of Ganymede

A. Migliorini[1,*], Z. Kanuchova[2], S. Ioppolo[3,*], M. Barbieri[4], N.C. Jones[5], S.V. Hoffmann[5], G. Strazzulla[6], F. Tosi[1], G. Piccioni[1]

[1] Institute of Space Astrophysics and Planetology, IAPS-INAF, Rome 00133, Italy
[2] Astronomical Institute of Slovak Academy of Sciences, Tatranská Lomnica 059 60, Slovakia
[3] School of Electronic Engineering and Computer Science, Queen Mary University of London, London E1 4NS, UK
[4] Instituto de Astronomía y Ciencias Planetarias, INCT, Universidad de Atacama, Copiapó, Chile
[5] ISA, Department of Physics and Astronomy, Aarhus University, 8000 Aarhus C, Denmark
[6] Osservatorio Astrofisico di Catania, INAF, Catania 95123, Italy

[*] Corresponding author.
E-mail addresses: alessandra.migliorini@inaf.it (A. Migliorini), s.ioppolo@qmul.ac.uk (S. Ioppolo)

Keywords: Cosmochemistry - Ultraviolet observations - Ices, UV spectroscopy - Jupiter, satellites - Satellites, surfaces

**ABSTRACT**
Since its first identification on the surface of Ganymede in 1995, molecular oxygen ($O_2$) ice has been at the center of a scientific debate as the surface temperature of the Jovian moon is on average well above the freezing point of $O_2$. Laboratory evidence suggested that solid $O_2$ may either exist in a cold (<50 K) subsurface layer of the icy surface of Ganymede, or it is in an atmospheric haze of the moon. Alternatively, $O_2$ is constantly replenished at the surface through ion irradiation of water-containing ices. A conclusive answer on the existence of solid $O_2$ on the surface of Ganymede is hampered by the lack of detailed, extensive observational datasets. We present new ground-based, high-resolution spectroscopic observations of Ganymede's surface obtained at the Telescopio Nazionale Galileo. These are combined with dedicated laboratory measurements of ultraviolet-visible (UV-vis) photoabsorption spectra of $O_2$ ice, both pure and mixed with other species of potential interest for the Galilean satellites. Our study confirms that the two bands identified in the visible spectra of Ganymede's surface are due to the (1,0) and (0,0) transition bands of $O_2$ ice. Oxygen-rich ice mixtures including water ($H_2O$) and carbon dioxide ($CO_2$) can reproduce observational reflectance data of the Ganymede's surface better than pure $O_2$ ice in the temperature range 20-35 K. Solid $H_2O$ and $CO_2$ also provide an environment where $O_2$ ice can be trapped at higher temperatures than its pure ice desorption under vacuum space conditions. Our experiments at different temperatures show also that the (1,0)/(0,0) ratio in case of the $CO_2:O_2$ = 1:2 ice mixture at 35 K has the closest value to observations, while at 30 K the (1,0)/(0,0) ratio seems to be mixture independent with the exception of the $N_2:O_2$ = 1:2 ice mixture. The present work will be part of a laboratory spectroscopic database in support of the ESA/JUICE mission to the Jovian system.

## 1. Introduction

Most of the moons of the Solar System have exospheres that are usually tenuous atmospheres consisting of volatile molecules, such as molecular nitrogen, molecular oxygen, carbon monoxide, ozone, carbon dioxide, water, methane ($N_2$, $O_2$, CO, $O_3$, $CO_2$, $H_2O$, $CH_4$). Other molecules, including S-bearing species observed on Io, hydrocarbons, and NaCl, that are volcanically emitted or created when surface materials undergo irradiation by the solar wind or ions accelerated by a nearby giant planet's magnetic field, may produce collisional exospheres, as in the case of Ganymede, or surface-bound exospheres (Hall et al., 1995, 1998; Lellouch et al., 2003; Liang et al., 2005; Teolis and Waite, 2016; Honniball et al., 2021). In most cases, the surface temperature of the moons is low enough to allow for the presence of ice layers that can reach thicknesses of several kilometers. Water ice dominates the surfaces of the icy Galilean satellites: Europa, Ganymede, and Callisto, as well as those of Saturnian moons. On the Ganymede's surface, the daytime temperatures range from 90 to 152 K (Orton et al., 1996), and are significantly lower at night-time. Hence, more volatile species such as $N_2$, CO, and $CH_4$, in several instances, dominate the satellites' surface composition together with crystalline and amorphous $H_2O$ ice, ammonia hydrates ($NH_3 \cdot H_2O$), and dark organic material (e.g., Cruikshank et al., 2020).

To date, molecular oxygen, a volatile species relevant to life on Earth, has been detected on several Galilean and Saturnian moons such as Callisto, Europa, Ganymede, Dione, and Rhea as well as in the coma of comet 67P/Churyumov-Gerasimenko (Spencer et al., 1995; Hall et al., 1995, 1998; Carlson et al., 1999; Spencer and Calvin, 2002; Liang et al., 2005; Bieler et al., 2015; Teolis and Waite, 2016). Observational evidence suggests that the $O_2$ abundance in comet 67P/CG is compatible with interstellar values (Bieler et al., 2015; Heritier et al., 2018). It is relevant to say that some of these observations are relative to solid state features, some are relative to the gaseous phase, some are direct measurements, while others are inferred. In the case of $O_2$ on the surface of Jovian and Saturnian moons, since the surface temperature of these objects is above the expected desorption temperature of pure $O_2$, the exact process linked to its formation and preservation in the solid phase is still under debate.

For instance, if present in the solid phase, $O_2$ must be constantly replenished through radiolysis to compensate for its depletion due to desorption (Brown et al., 1982; Baragiola and Bahr 1998; Baragiola et al., 1999). If $O_2$ has a primordial origin, it could be efficiently trapped and preserved in the deeper layers of water-rich ice (Collings et al., 2004). Regardless of its initial origin, if present in the solid phase, subsurface $O_2$ could survive under a very slow balance of replenishment and loss due to surface temperatures above its thermal desorption. Vidal et al. (1997) suggested that solid $O_2$ may exist in a cold (<50 K) subsurface layer or in an atmospheric haze of Ganymede. Teolis et al. (2006) found that ion irradiation (100 keV $H^+$) during the deposition of $H_2O$ at 130 K causes the burial of a high concentration of radiolytic $O_2$ from which solid ozone is formed. In their experiments, Teolis et al. (2006) infer the presence of solid $O_2$ from the appearance of the $O_3$ band in the UV spectrum of irradiated samples. Moreover, the presence of $O_2$ is confirmed by means of mass spectrometry. The authors concluded that the enhanced trapping of oxygen in surface ices depends on the surface temperature and should vary locally, depending on the rates of irradiation and recondensation. Clearly, more work is needed to better understand the presence of $O_2$ on the surface of some inner Solar System moons.

Ultraviolet-visible (UV-vis) spectroscopy has been extensively used to identify $O_2$ in the Solar System. In 1995, ground-based observations revealed two features centered at 577.3 and 627.5 nm that were attributed to solid $O_2$ in Ganymede's trailing hemisphere (Spencer et al., 1995). However, no evidence of the same features was found in the leading side of Ganymede and it was shown that the bands were only visible for orbital longitudes between 200° and 330°. More recently, spectra with a higher signal-to-noise ratio (SNR) of the surface of Europa and Callisto have been acquired, revealing the presence of an absorption band at 577.3 nm weaker than the one found on Ganymede (Spencer and Calvin, 2002). The band depth relative to the spectral continuum reported for the Europa leading and trailing sides was 0.3%, while it was ten times stronger, i.e., 3%, on Ganymede's trailing face. During that observing campaign, the authors also monitored the leading face of Europa, which showed an $O_2$ feature similar to the one observed in the trailing face of the satellite. Unfortunately, the Visual and Infrared Mapping Spectrometer (VIMS) on board the Cassini mission was not able to resolve the solid $O_2$ absorption bands in the Galilean satellites that were observed during the Millennium flyby, which occurred at the end of 2000 (McCord et al., 2004). Further searches for the 577.3 and 627.5 nm features on the saturnian satellites have only yielded upper limits (Spencer, 1998). Despite the numerous investigations of the $O_2$ bands using ground- and space-based observations, there is still a lack of agreement in the band positions and their relative intensities. A possible explanation lies on the fact that the available measurements present a largely variable spatial resolution, which might be important in the case of a heterogeneous surface like in the case of Ganymede. Hence, more detailed observational evidence for the presence of solid $O_2$ on the surface of the Galilean moons is needed.

In the vacuum ultraviolet (VUV) and ultraviolet-visible (UV-vis) spectral ranges (≤200 nm and 200-700 nm, respectively), solid $O_2$ presents several absorption features such as the stronger Schumann-Runge band at 140 nm and the aforementioned two weaker absorptions at 577.3 and 627.5 nm due to a double electronic transition of adjacent $O_2$ molecules (e.g., Landau et al., 1962; Mason et al., 2006). Laboratory experiments showed that the spectral profile and peak positions of the two UV-vis $O_2$ absorption bands can be altered when $O_2$ is embedded in different ice environments (Landau et al., 1962). The shape of the two absorption bands can also reveal the crystalline or amorphous phase of solid $O_2$ in an ice matrix and hence it can provide hints on the local surface temperature of the ice (Vidal et al., 1997). A comprehensive summary of $O_2$ properties in the visible and near-infrared (VNIR) is provided in Copper et al. (2003). A systematic laboratory study of the spectral profiles and peak positions of the two UV-vis bands of $O_2$, pure and mixed with other species, as a function of the ice temperature and with spectral resolution that matches observations can aid a better interpretation of observational data of the surfaces of Jupiter's moons.

In this work, we present new spectroscopic high-resolution observational data of Ganymede, obtained with the High Accuracy Radial velocity Planet Searcher (HARPS-North) spectrograph at the Telescopio Nazionale Galileo (TNG), which is a 3.58 m telescope located at the top of the Roque de Los Muchachos Observatory in the Canarian island of La Palma, to investigate peak positions and profiles of the two $O_2$ spectral features in the optical range. Supporting laboratory experiments were performed to measure the UV-vis spectra of frozen $O_2$, pure and mixed with other relevant molecules ($H_2O$, $CO_2$, and $N_2$). Laboratory measurements were obtained at the ASTRID2 synchrotron light source at Aarhus University in Denmark. Experimental data will be included in a VUV-UV-vis ice database (Banks, 2012) in support of the ESA space mission JUICE (JUpiter ICy moons Explorer), set to make detailed observations of the giant gaseous planet Jupiter and three of its largest moons, Ganymede, Europa, and Callisto (Grasset et al., 2013). Onboard JUICE, the MAJIS (Moons And Jupiter Imaging Spectrometer) instrument is an imaging spectrometer composed of two channels: a VIS-NIR channel covering the 0.5-2.35 μm range, and an IR channel covering the 2.25-5.54 μm range. Both ranges are sampled in 508 spectral channels with an average spectral sampling value of 3.66 and 6.51 nm/band, respectively (Piccioni et al., 2019). The typical output of MAJIS is a hyperspectral image, i.e., a scan of the observed scene, where each pixel has a spectrum potentially covering the entire 0.5-5.5 µm range. The MAJIS primary objective is the hyperspectral mapping of the surfaces of the three aforementioned icy Galilean satellites.

At Ganymede, around which JUICE will enter orbit towards the end of its mission, MAJIS is committed to achieve broadly regional coverage with a spatial resolution varying between 1 and 5 km/px (3 km/px on average), which roughly corresponds to the best spatial resolution previously achieved by the Galileo/NIMS instrument only on a few specific locations. Prior to the Ganymede orbit phase, the JUICE mission profile foresees eight flybys of Ganymede and a larger number of Callisto flybys (10 to 20, depending on the final mission profile) with variable minimum altitudes over the surface with closest approach values being 400 km and 200 km, respectively, plus two Europa flybys as close as ~400 km. Several features are expected to be identified on the surface of Ganymede and Europa, through spectral investigation in the VNIR for both. Although the expected magnitude of the UV-vis $O_2$ bands presented here are at the limit of noise with MAJIS (Noise Equivalent Spectral Radiance, or NESR, of 0.032 and 0.02 W m$^{-2}$ sr$^{-1}$ µm$^{-1}$ calculated at 0.5764 and 0.6274 µm, respectively), they might be detected, if present, by averaging tens of spectra, allowing for a better future characterisation of the chemical composition of the surface of Jupiter's moons.

## 2. Ground-based observations

High-resolution spectroscopy of Ganymede acquired with the HARPS-North spectrograph at the Telescopio Nazionale Galileo, were selected from the TNG archive to investigate the presence of $O_2$ absorption bands on Ganymede's surface. For calibration purposes, B-type star 44 Gem (B8V) was also observed during the night of January 4$^{th}$ (airmass 1.12) with the same instrument setup, while solar analog star HD 221627 (G1IV) was observed during the night of January 5$^{th}$ (airmass 1.11). Ganymede was observed on January 5$^{th}$, 2014, during the observations of the Earth transit across the Sun as seen from Jupiter's moons (Molaro et al., 2015), and 188 spectra have been acquired with airmass ranging from 1.01 to 2.6. For reducing the effect of airmass variation, we restricted our analysis only to the first 120 spectra of Ganymede, that were gathered with airmass less than 1.17. On the observing date, Ganymede had a visual magnitude of 4.6; the commanded exposure time was ranging between 60 s and 300 s, sufficient to have a SNR of about 200. We use the same dataset reduced by Molaro et al. (2015), where the spectra were processed with the HARPS-North pipeline and the resulting spectra were corrected for the relative motion of the observer and the satellites.

To avoid any contamination of the H-alpha at 656.3 nm, which is strong in the spectrum of the standard star 44 Gem, we consider only the region 550.0-645.0 nm. This is required to prevent an over correction of the spectrum beyond 656 nm. This portion of the spectrum is fitted with a 3-deg polynomial spline, and used to normalize all the spectra. Since the telluric star 44 Gem was not observed on the same night as of Ganymede and HD 221627, a special procedure was adopted to remove all contributions of telluric lines that could undermine the detection of $O_2$ features. The spectra of HD221627 alone could not be used for removal of both solar and telluric lines, since the star has a radial velocity of -5.7 km/s, and hence shifting this spectrum by the corresponding Radial Velocity (RV) to match the Ganymede spectra will shift rigidly also the telluric lines and the division of both spectra will not remove these features. The adopted procedure included the removing of telluric contribution from the solar analog spectrum, the correction of all Ganymede's spectra for solar contribution, correction of the Ganymede spectra of previous step for telluric contribution to finally produce a mean Ganymede spectrum. First we corrected the spectra of HD 221627 for telluric absorption using the spectra of 44 Gem, calculating the Cross Correlation Function (CCF) between the two spectra, shifting the spectra for the corresponding value and dividing the spectra of HD 221627 by the spectra of 44 Gem taking into account also the difference in airmass using the formula (1):

$$\mathrm{HD}_{telluric\_corrected} = \frac{\mathrm{HD}\ 221627}{44\ \mathrm{Gem}^{(airmassHD/airmass44Gem)}} \qquad (1).$$

For each Ganymede spectra, we calculate the CCF against the spectrum of HD221627 obtained in the previous step, we corrected for the RV value, and divided each Ganymede spectra by the spectrum of HD221627 (using the airmass correction). The resulting spectra of Ganymede are corrected for solar lines; however, they still show some telluric contribution. For these spectra of Ganymede, we calculate the CCF against the spectrum of 44 Gem, corrected for RV and divided by the spectra of 44 Gem. Finally, all these Ganymede spectra were shifted in RV respect to the first Ganymede observation in the series, and all the spectra was averaged and smoothed using a 500 points boxcar.

The final spectrum is shown in panel (a) of Figure 1. At these observing conditions, the obtained spectrum refers to a sub-observer longitude in the range 215.4-220.7 W and a sub-observer latitude of 1.78 N, as derived by using the SPICE toolkit (Acton, 1996). Hence, the TNG measurements were pointing at the trailing hemisphere of Ganymede, in the anti-Jovian side. In Figure 1, a band centered at about 576.1 nm is observed in the spectrum of Ganymede and a second band centered at about 624.0 nm is also visible. Weaker absorption bands in the latter spectral range are due to a sub-optimal removal of telluric bands that affects the 624.0 nm band profile and prevent an accurate determination of the band peak position. It is worth noticing the presence of the Na absorption bands at 589 nm, which are probably a consequence of the fact that the solar analog we are using is not perfect. In the TNG measurements, the band positions were obtained by deriving the intercept of the left- and right-hand curve fitting the two bands. An error of ±0.5 nm is associated with the identification of the 576.1 band position, which includes the uncertainty due to data reduction and curve fitting. A higher error (±5 nm) is expected for the 624.0 nm band due to telluric noise. Overall, our TNG spectra of Ganymede's surface are in agreement with previous observational data presented in Spencer et al. (1995; Fig. 1 panel b), where a Ganymede/Callisto ratio spectrum was presented with two bands visible at 577.3 and 627.5 nm. A different observational methodology and analysis between our measurements and those from the literature can likely explain minor discrepancies in terms of peak positions and band profiles. Due to the telluric noise of the 624.0 nm band in the TNG measurements, in the next sections we compare our laboratory results to the data from Spencer et al. (1995). For completeness, Appendix A presents a comparison between our TNG data and laboratory work.

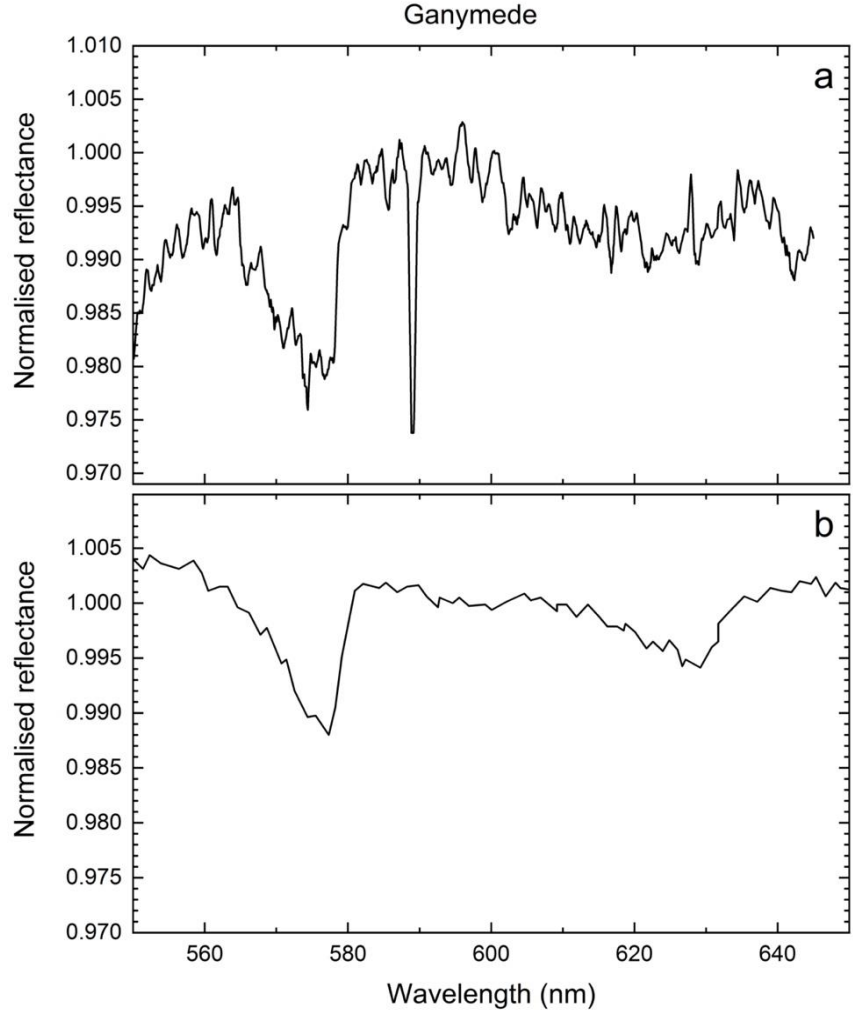

**Fig. 1.** Observational spectrum of Ganymede in the 550-650 nm range. Panel (a) shows the high-resolution spectrum of Ganymede obtained at the TNG observational facility using the HARPS-North high-resolution spectrograph. The Na absorption band at 589 nm is due to the solar analog. Panel (b) presents observations of Ganymede by Spencer et al. (1995) for comparison.

### 3. Synchrotron measurements

To aid the interpretation of observational data, laboratory measurements were performed at the ISA ASTRID2 storage ring, Aarhus University, Denmark. First to become operational on the new 3$^{rd}$ generation light source ASTRID2, the AU-UV beam line used in the present work can accommodate a large variety of end-station experimental setups to measure photoabsorption spectra in the range 115 to 700 nm by using two gratings (Eden et al., 2006; Palmer et al., 2015). The high energy grating (HEG) settings used in previous work on VUV photoabsorption spectroscopy of $O_2$-rich ices in the range 120 to 340 nm corresponded to a typical VUV beam flux of $10^{10}$ photons/s/100 mA and a wavelength step size between 0.05 and 0.2 nm was used, depending on the width of the spectral features to be resolved (Ioppolo et al., 2020). Here, to access the UV-vis spectral range, we used the low energy grating (LEG) with spectra acquired in transmission-absorption mode over the wavelength range of 310-670 nm with a 0.2 nm wavelength step size. The typical photon flux per point was in the range of 1 to 2.5x$10^{10}$ photons/s/100 mA and collection time per point was 2 s.

The experiments described here were performed using the custom-made Portable Astrochemistry Chamber (PAC), a high vacuum (HV) system with a base pressure of $10^{-9}$ mbar. The apparatus is described in detail in Ioppolo et al. (2020, 2021). Here briefly, the PAC consists of a compact spherical cube chamber (Kimball Physics) connected to a turbo molecular pump, a closed cycle helium cryostat with a base temperature of 20 K, and a 1 keV electron gun (Kimball Physics). At the center of the main chamber, a substrate magnesium fluoride window ($MgF_2$, Crystran) is mounted in a sample holder made of oxygen-free high conductivity copper (OFHC, Goodfellow) in thermal contact with the cryostat. The temperature of the substrate is measured with a silicon diode (DT-670, Lakeshore) and can be controlled in the range 20-300 K by means of a Kapton tape heater (Omega) connected to the OFHC block and regulated with a temperature controller system (Oxford Instruments). The chamber is enclosed with $MgF_2$ windows (Crystran). A differentially pumped rotary stage mounted in between the main chamber and the cryostat head is used to change the position of the bare substrate with respect to the incoming UV photons, dosing line inlet, and electron gun output. Ice samples were prepared by deposition of pure $O_2$ and mixtures of $H_2O:O_2$, $N_2:O_2$, and $CO_2:O_2$ on the $MgF_2$ substrate window cooled down to 20 K. Gas mixtures were made in a dedicated pre-chamber and admitted into the main chamber by an all-metal needle valve. The ratios of the mixing gases were determined via partial pressures measured by means of a mass-independent capacitance manometer prior to deposition. Direct deposition was carried-out at normal incidence with respect to the substrate through a 6 mm diameter tube extending inside the main chamber. For all studied ices, ice thicknesses were monitored by a well-established HeNe laser technique (Born and Wolf, 1970; Goodman, 1978; Baratta and Palumbo, 1998) and were initially in the range of 7±0.5 µm with a deposition rate of about 0.5 µm /min. Due to the low absorption cross section of the two $O_2$ bands, depositions of 7 µm thick ices were needed to achieve an SNR of at least 5:1. During UV-vis spectrum acquisition, the sample was positioned with its surface normal to the incident synchrotron light. The intensity of the transmitted synchrotron light was measured by a photodiode (AXUV100G, Opto Diode). Thermal processes carried-out during the experiments resulted in a decrease of the initial ice thickness visible in the spectra at temperatures above 20 K (see Fig. 2). However, a quantitative estimate of the film thicknesses for spectra at all temperatures was not possible because of the lack of further measuring devices beyond the HeNe laser technique used during deposition only, such as a quartz crystal microbalance.

4. **Experimental Results**

Figure 2 shows laboratory UV-vis photoabsorption spectra of pure $O_2$ deposited at 20 K and heated to 30, 35, and 40 K. In this spectral range, two $O_2$ bands are clearly visible. Under our HV experimental condition, at temperatures higher than about 35 K, pure $O_2$ sublimates and the bands disappear, in agreement with measurements reported in Baragiola and Bahr (1998). The two electronic absorption features are due to the $a^1\Delta_g + a^1\Delta_g - X^3\Sigma^-_g + X^3\Sigma^-_g$ transition, where the (0,0) band is centered at 626.2 nm and the (1,0) at 575.9 nm at 20 K. In Figure 2, two vertical dashed lines indicate the peak positions of the $O_2$ bands observed on Ganymede by Spencer et al. (1995). The peak positions of the pure $O_2$ (1,0) features in the 35 K spectrum seems to qualitatively match observational data better than other temperatures, showing also a similar relative intensity for the two transitions. The laboratory spectral profiles for the (0,0) and (1,0) bands confirm that oxygen deposited at 20 K is in its α form (Landau et al., 1962). However, when heated to 30 K and 35 K, the shape of the two bands changes. The right side of the (1,0) band at 576 nm becomes less steep, while the (0,0) band is redshifted. This is indicative of a changing in the structure of $O_2$ ice that is consistent with the β form of solid $O_2$ as assigned by Landau et al. (1962). In a previous study that used the PAC apparatus at ASTRID2 to deposit pure $O_2$ ice at 20 K under similar experimental conditions of those presented here, we reported the full desorption of molecular oxygen at 35 K (Ioppolo et al., 2020). It is known that under vacuum conditions molecular oxygen desorbs at temperature above 30 K (Acharyya et al., 2007). Hence, any desorption temperature discrepancy between our two studies can be attributed to the difference in ice thickness, being 0.25 µm in Ioppolo et al. (2020) and ~ 7 µm here. A thicker ice requires a longer time to achieve total desorption, allowing us to record UV-vis spectra of the pure $O_2$ ice at 35 K prior to its full desorption. The 40 K spectrum of Fig. 2 is flat, indicating the complete desorption of the ~ 7 µm thick $O_2$ ice. Landau et al. (1962) describe a γ phase for $O_2$ at 49 K under their experimental conditions. In agreement with our TNG observations and Spencer et al. (1995), unlike the α and β forms, the γ phase presents a weaker (0,0) band compared to the (0,1). By qualitatively comparing the peak intensities and positions of the three forms of pure $O_2$ ice (Landau et al., 1962) with the available observational data, one can infer that $O_2$ ice on Ganymede is a combination of the different ice forms. This finding is in agreement with the hypothesis that $O_2$ is trapped under Ganymede's ice surface in a cold (<50 K) subsurface layer (Vidal et al. 1997). To verify this idea, we have performed further laboratory experiments on mixed $O_2$-containing ices in the UV-vis spectral range to investigate any spectral profile change of the two $O_2$ bands as a function of the mixing ratio, ice composition, and temperature.

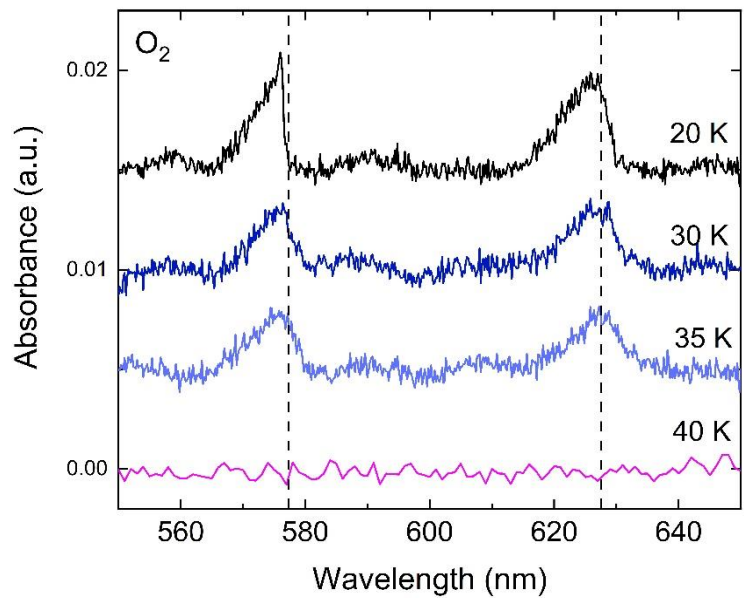

**Fig. 2.** UV-vis photoabsorption spectra (550-650 nm) of pure molecular oxygen, deposited at 20 K (black curve) and heated to 30 K (dark blue), 35 K (light blue), and 40 K (magenta). All spectra were acquired with a wavelength step size of 0.2 nm, except for the 40 K spectrum that has a step size of 1 nm. For comparison, two vertical dashed lines are centered on the peak positions of the two $O_2$ bands (577.3 nm and 627.5 nm) as observed on Ganymede by Spencer et al. (1995).

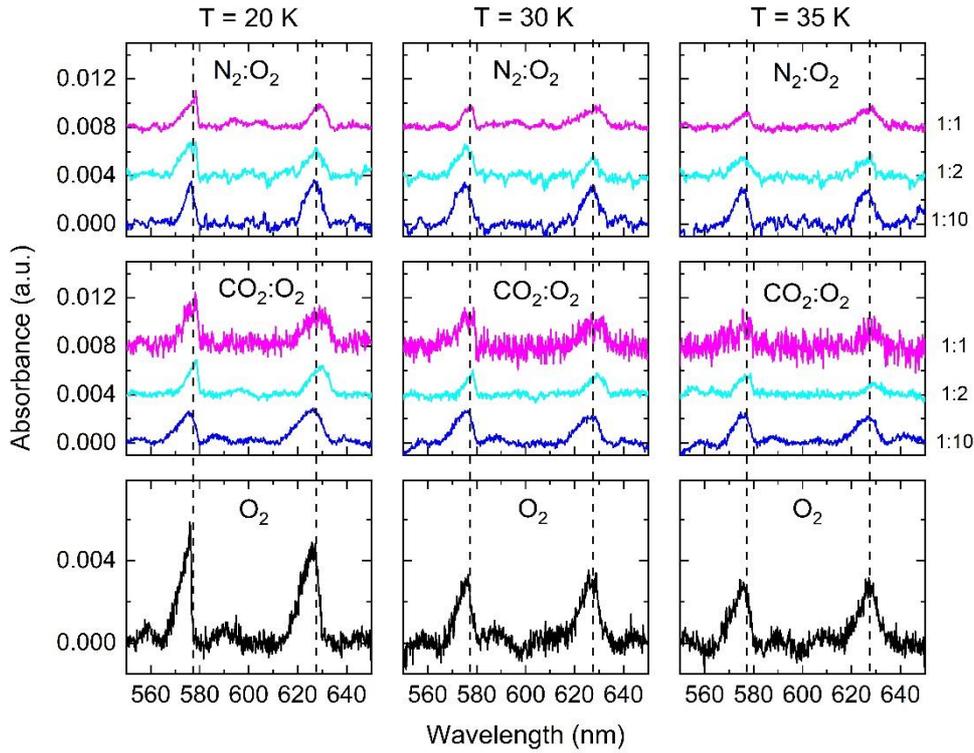

**Fig. 3.** UV-vis photoabsorption spectra (550-650 nm) of pure oxygen (black lines, bottom panels) compared to molecular oxygen mixed 1:1 (magenta lines), 1:2 (light blue lines), and 1:10 (dark blue lines) with $CO_2$ (central panels) and $N_2$ (top panels) deposited at 20 K (left panels) and heated to 30 K (middle panels) and 35 K (right panels). Vertical dashed lines indicate the position of the bands observed in the Ganymede's spectrum by Spencer et al. (1995).

Figure 3 compares laboratory UV-vis photoabsorption spectra of pure $O_2$ ice (black lines, bottom panels) acquired at different temperatures (20, 30, and 35 K) in the spectral range 550-650 nm with spectra of molecular oxygen mixed with other species with different ratios (X:$O_2$ = 1:1, 1:2, and 1:10, with X = $CO_2$ and $N_2$) deposited at 20 K and heated to 30 and 35 K. The spectral profile, shape, relative intensities, and peak position of the two $O_2$ bands clearly change as a function of the mixing ratios, ice composition, and temperature. Vertical lines in Fig. 3 indicate the position of the bands observed in the Ganymede's spectrum shown in Spencer et al. (1995). As for the case of pure $O_2$, spectra at higher temperatures show band positions that are closer to those observed on Ganymede by Spencer et al. (1995). Moreover, the presence of species in the ice other than $O_2$ seems to further improve the comparison to observations and an overall better agreement with the observational band peak positions is qualitatively obtained in the case of the 1:10 mixtures, that is when $O_2$ is the main ice component. The (0,0) and (1,0) spectral profiles are generally asymmetric, therefore, to characterize them and identify peak positions more accurately, we have fitted all available spectra with the asymmetric double sigmoidal function below:

$$y = y_0 + A \cdot \frac{1}{1+e^{-\frac{x-x_c+\frac{w_1}{2}}{w_2}}} \cdot \left(1 - \frac{1}{1+e^{-\frac{x-x_c-\frac{w_1}{2}}{w_3}}}\right) \qquad (2),$$

where $y_0$ is an offset, $A$ is the amplitude of the asymmetric double sigmoidal function, $x_c$ is its center, $w_1$ is the full width of half maximum, $w_2$ is the variance of the low-energy side of the curve (higher wavelengths with respect to $x_c$), and $w_3$ is the variance of the high-energy side of the curve (lower wavelengths with respect to $x_c$).

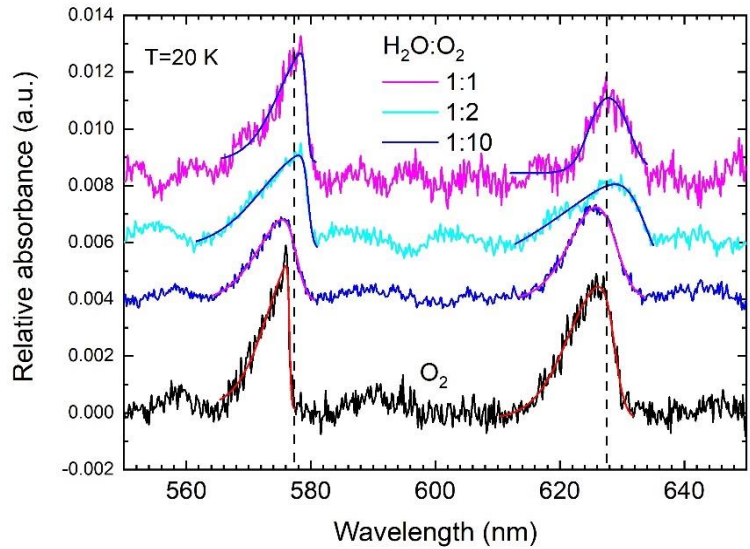

**Fig. 4.** UV-vis photoabsorption spectra (550-650 nm) of pure oxygen (black line) compared to molecular oxygen mixed 1:1 (magenta line), 1:2 (light blue line), and 1:10 (dark blue line) with $H_2O$ deposited at 20 K. The asymmetric double sigmoidal fits of the UV-vis photoabsorption bands (solid red for pure $O_2$, dark blue for $H_2O:O_2$ = 1:1 and 1:2, and magenta for the 1:10 mixture) are shown.

Figure 4 shows selected fits of laboratory data with Eq. (2). The figure presents a photoabsorption spectrum of pure $O_2$ at 20 K (black line) compared to spectra of $H_2O:O_2$ = 1:1, 1:2, and 1:10 at 20 K (magenta, light and dark blue lines, respectively). Vertical dashed lines represent observational peak positions for the (0,0) and (1,0) bands found in their observations of Ganymede by Spencer et al. (1995). In the case of $H_2O:O_2$ mixture, band peak positions seem to be shifted toward longer wavelengths compared to the case of pure $O_2$ ice. This is more evident for the (1,0) transition and when $H_2O$ is in a larger percentage in the mixture, as also reported in previous works by Baragiola and Bahr (1998). The authors studied the wavelength shift with temperature for $H_2O-O_2$ mixtures, reporting a redshift with increasing temperature. In addition, it can be seen that the 1:1 and 1:2 concentrations of the $H_2O:O_2$ mixture show a stronger (1,0) transition in absorption than the (0,0) transition, while the two bands have a similar height in the 1:10 mixture, where $O_2$ is the dominant species. This result is in agreement with previous laboratory measurements in Vidal et al. (1997). Although it has been suggested that molecular oxygen is mixed and/or layered with water ice on the surface of the Galilean moons, our selected mixtures at 20 K do not generally improve the match of the laboratory and observational peak positions compare to that of pure molecular oxygen. However, it should be noted that we do not have laboratory data of mixed $H_2O:O_2$ ices at 30 and 35 K. As for the case of other $O_2$-rich ice mixtures, it is reasonable to expect that the match with observations would improve at higher temperature. Future laboratory work is foreseen to extend our VUV-UV-vis ice database with spectra of ice mixtures containing more than two ice components, different ratios, and temperatures. Nevertheless, the available dataset can already be used to retrieve valuable information that will guide future observations and laboratory work.

In Figure 5, the positions of the (0,0) and (1,0) bands maxima as obtained by the fit between laboratory data and Eq. (2) are plotted at different temperatures (20 K left panel, 30 K middle panel, and 35 K right panel). Here, peak positions of the two bands observed on Ganymede in Spencer et al. (1995; black cross) are presented for comparison. A close comparison between observations and laboratory data indicates that at 20 K pure molecular oxygen cannot qualitatively reproduce the peak position of the (1,0) and (0,0) transition bands. The observed peak positions are only reproduced when molecular oxygen is mixed with other species. Traces of molecular nitrogen in an oxygen-rich ice ($N_2:O_2$ = 1:10) seems to reproduce the peak positions of both bands as observed by Spencer et al. (1995). Figure 5 suggests that mixtures of water and carbon dioxide in oxygen-rich ices at 20 K with mixing ratios between 0.5 and 0.1 can qualitatively match the observed band peak positions in the case of Spencer et al. (1995). At 30 K, several ice mixtures as well as pure $O_2$ ice can reproduce the peak position of the (0,0) transition band only, while other mixtures can exclusively reproduce the (1,0) band. At 35 K, all selected laboratory values of the band peak positions seem to converge with observations within experimental and observational uncertainties better than at other temperatures. This is an indication that the observed (1,0) and (0,0) transition bands are due to solid molecular oxygen trapped in a more refractory ice at temperatures equal or higher than 35 K, as expected on the surface of Ganymede.

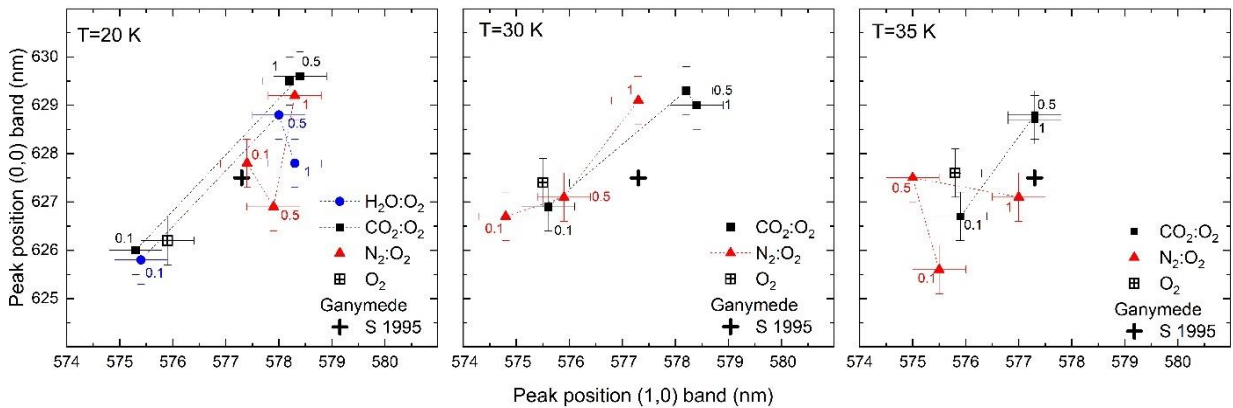

**Fig. 5.** Peak position of (1,0) and (0,0) transition bands of pure $O_2$ (black boxed cross) and $O_2$ in various ice mixtures (with $H_2O$ blue circle, $CO_2$ black square, and $N_2$ red triangle) deposited at 20 K compared to peak positions of relevant bands as observed in the spectra of Ganymede in Spencer et al. (1995; black cross). Peak positions of laboratory data are obtained from the data fit with a double sigmoidal function (Eq. (2)). Error bars highlight uncertainties in the determination of peak positions

due to spectral noise. Numbers at data points represent the X/$O_2$ mixing ratios, where X is $H_2O$, $CO_2$, and $N_2$.

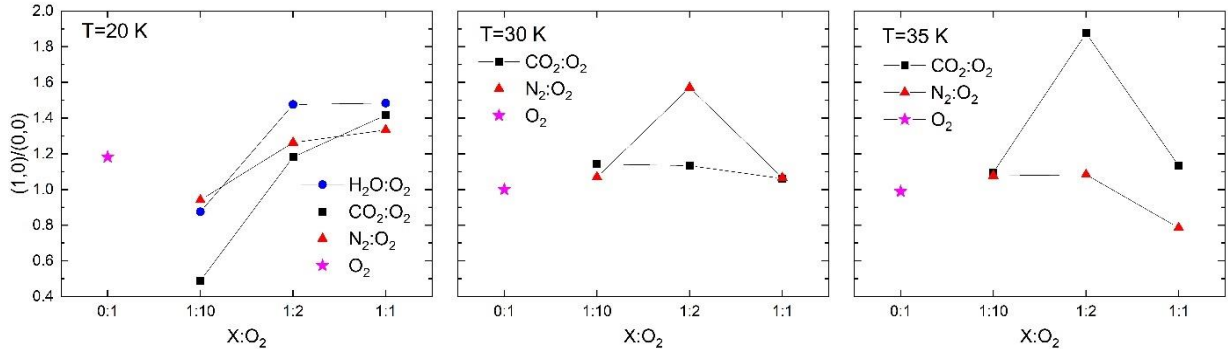

**Fig. 6.** Ratio of the intensities in absorbance of the (1,0) and (0,0) $O_2$ transition bands at their peak positions ((1,0)/(0,0)) as a function of the X:$O_2$ ice mixing ratios, where X is $H_2O$, $CO_2$, and $N_2$ (blue circle, black square, and red triangle, respectively). The (1,0)/(0,0) ratio for pure $O_2$ ice (magenta star) is also shown for comparison. Left, middle, and right panels show available data acquired at 20, 30, and 35 K, respectively. The observational analog (1,0)/(0,0) ratio from Spencer et al. (1995) is approximately 2 in value and is represented by the top-horizontal axis of all panels.

In Figure 6, the ratio of the intensities in absorbance of the (1,0) and (0,0) $O_2$ transition bands at their peak positions ((1,0)/(0,0)) as obtained by the fit between laboratory data and Eq. (2) are plotted at different temperatures (20 K left panel, 30 K middle panel, and 35 K right panel) and as a function of the X:$O_2$ ice mixing ratios, where X is $H_2O$, $CO_2$, and $N_2$ (blue circle, black square, and red triangle, respectively). The (1,0)/(0,0) ratio for pure $O_2$ ice (magenta star) is also shown for comparison. The value of the corresponding (1,0)/(0,0) ratio of the two bands observed on Ganymede by Spencer et al. (1995) is approximately 2 and is represented by the top-horizontal axis of all panels. All ratios obtained experimentally are lower than the observational one because the observed (0,0) transition band is weaker than the (1,0) at their peak positions compared to our laboratory data (see Fig. 1). Experimental ratios obtained from the mixtures at 20 K seem to indicate that the 1:1 mixtures present weaker (0,0) transitions relative to the (1,0). Hence, ratios are closer to the observed one, especially in the case of the $H_2O$:$O_2$ = 1:1 ice mixture. At 20 K, the 1:10 mixtures present the lowest (1,0)/(0,0) ratio compared to all other cases, that is the worse match with observations. At 30 K, apart for the $N_2$:$O_2$ = 1:2 ice mixture, the (1,0)/(0,0) ratio seems to be mixture independent. At 35 K, the $CO_2$:$O_2$ = 1:2 ice mixture presents the closest (1,0)/(0,0) value to the observed one compared to all other mixing ratios and temperatures considered in this work.

Figure 7 compares the observed spectrum of Ganymede in reflectance obtained by Spencer et al. (1995) with some of the laboratory data converted to transmittance and selected according to the results shown in Figs. 5 and 6 to determine whether the observed band profiles can be reproduced by a set of laboratory data. To determine how well the Ganymede band is reproduced by a laboratory spectrum (in terms of band profile and peak position), we have calculated residuals - the difference between the laboratory and observational spectra, and the sum of squared residuals (SSR) reported in the residual plot of each panel in Fig. 7. A direct quantitative indicator of the best match of observational and laboratory spectra would be the smallest SSR value. However, all SSR values are comparably small and, because they are calculated over the whole studied spectral range, the interpretation of the results could be misleading. A simple qualitative inspection of the residual spectrum, particularly in the regions of two $O_2$ bands can be helpful: a flat residual spectrum close to zero indicates the match of observed and experimental data. Figure 7 is prepared such that all panels on the left of the figure present the best matches at 20, 30, and 35 K for all mixtures compared to pure $O_2$ ice, bottom panels. The right panels of the figure show the 1:10 mixtures of the corresponding $X:O_2$ ices. Pure $O_2$ ice at 35 K can better reproduce the observed (1,0) transition band with a flat residual spectrum close to zero compared to pure $O_2$ at other temperatures. However, the (0,0) band is not reproduced satisfactorily at the same time. Laboratory spectra of mixed ices are needed to produce the minimal residual spectrum of both transitions of solid $O_2$. Particularly, the $H_2O:O_2 = 1:1$ mixture at 20 K, the $N_2:O_2 = 1:2$ mixture at 30 K, and the $CO_2:O_2 = 1:2$ mixture at 35 K (left top three panels of Fig. 7) are the laboratory components that best resemble at the same time peak position, intensity, and profile of the observed (1,0) and (0,0) bands by Spencer et al. (1995). The corresponding $X:O_2 = 1:10$ ice mixtures cannot reproduce the (0,0) transition satisfactorily (right top three panels of Fig. 7).

Our combined laboratory and observational work suggests that $O_2$ may be present on the surface of the satellite in patches surrounded by other molecular species and partially mixed with them. Water and carbon dioxide have the potential to efficiently trap oxygen in the ice, preserving it in the solid phase at temperatures higher than 50 K (e.g., Collings et al., 2004). Hence, they should be treated as important ice components, requiring further laboratory investigation in the UV-vis spectral range to consider different ratios and temperatures. Moreover, both $H_2O$ and $CO_2$ are molecules observed on the surface of the Galilean moons, making their mixtures with $O_2$ relevant to the surface of Ganymede and/or other objects in the Solar System. Future laboratory work on ternary mixtures containing $O_2$, $CO_2$, and $H_2O$ with different ratios are in the pipeline. Finally, although, at least in one of the considered cases, an $N_2:O_2$ mixture provided a good match with observations, we expect mixtures containing $N_2$ not to play a key role in trapping $O_2$ and, therefore, to be more relevant in the outer Solar System.

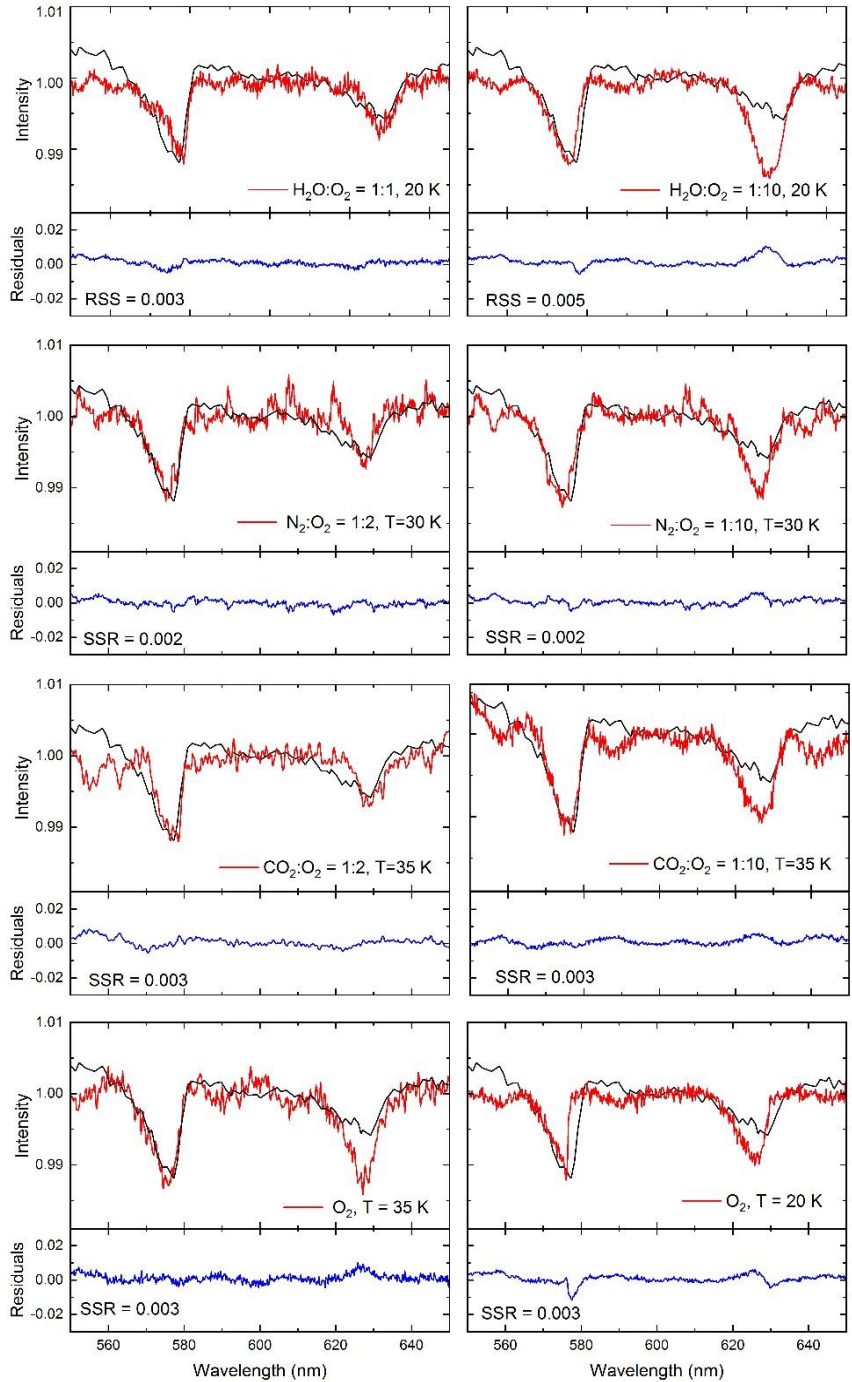

**Fig. 7.** Ground-based spectrum of Ganymede acquired in reflectance by Spencer et al. (1995) (black curve) compared to selected laboratory transmittance spectra of solid $O_2$, pure and mixed with $H_2O$, $CO_2$, and $N_2$, with different mixing ratios and ice temperatures (in red). Blue lines are the spectral residue that is the difference between black and red lines. SSR stands for sum of squared residuals.

## 5. Astrophysical Implications

At present solid $O_2$ has been firmly detected on Ganymede (on Europa and Callisto only one of the two bands here discussed has been observed) and only postulated in cometary comae (e.g., Heritier et al., 2018). It might also be present on many other objects in the outer Solar System and in Trans-

Neptunian Objects (TNOs). In many instances (e.g., Pluto), ices on the surfaces of those objects are very rich in $N_2$ (e.g., Cruikshank et al., 2020). Thus, it is possible that, if present, $O_2$ is mixed with nitrogen. Our experiments have been conducted at temperatures (20-50 K) that are typically below the average surface temperature measured on the Galilean satellites (Orton et al., 1996; Delitsky and Lane, 1998). This is because molecular oxygen is a volatile species that freezes at temperatures around 30 K under (ultra)high vacuum laboratory conditions. Hence, on the one hand, our laboratory results are more directly comparable with spectroscopic data of icy bodies in the outer Solar System than with Jovian moons. Yet, there is some evidence suggesting that even volatile species such as molecular oxygen can survive or be replenished on the surface of Jupiter's satellites such as Ganymede.

In the early 1980's, it was established that $O_2$ could be the product of ion bombardment on the Galilean moons (Johnson et al. 1983). Later on, Johnson and Jesser (1997) showed that $O_2$ could exist in fractured ice and bubbles at temperatures above the sublimation point. This would occur during the freeze-out of atmospheric $H_2O$ and $O_2$ onto the surface of Ganymede in regions unexposed to photons, i.e., at lower surface temperature (70-80 K; Delitsky and Lane, 1998). Other experiments on $H_2O$ ice films demonstrated that $O_2$ adsorption varies with temperature, showing a decrease with increasing ice temperature (Shi et al., 2009), as a consequence of the increase of $O_2$ desorption rate with temperature. However, the authors noted that the adsorbed $O_2$ is not permanently retained in the $H_2O$ film ice for temperatures higher than 80 K if no irradiation occurs. When irradiated, $O_2$ is preferentially trapped in the water film for temperatures up to 140 K, because irradiation has likely the function of closing the pores where oxygen is included. An additional set of experiments on $H_2O$ ice demonstrated that the amount of $O_2$ trapped in the water ice film was on the order of 10.4% of the total column density of the film itself at 40 K; this value decreased when the ice film was heated from 40 to 78 K (Shi et al., 2011). The presence of $O_2$ in such a large amount can account for the enhancement of production of other species, like $H_2O_2$ and $O_3$, and hence to the chemistry observed in the Ganymede case. Interestingly, Teolis et al. (2006) finds stable radiation-enhanced trapping of high radiolytic $O_2$ concentrations, on the order of tens of percent, in irradiated water ice during $H_2O$ deposition at high temperatures (130K), without co-depositing $O_2$. Unfortunately, Teolis et al. (2006) did not perform UV-Vis spectroscopy and only detected $O_2$ in the gas phase during the post-irradiation warm up of the sample.

Molecular oxygen could periodically cycle between the solid and the gas phase because of the day/night exposure of Ganymede's ice surface to photons from the Sun. Laboratory experiments by Collings et al. (2004) showed that the desorption temperature of oxygen is affected by the presence of amorphous water underneath or in a mixed layer, increasing to temperatures around 50-60 K. They also showed that a fraction of molecular oxygen can be trapped in a water-rich ice at temperatures exceeding the average surface temperature observed on Ganymede. Ion bombardment of water ice can supply molecular oxygen at the ice surface (Teolis et al., 2017). Moreover, apart from atmospheric $O_2$, implantation of $O_2^+$ ions is another possible source of $O_2$ in the ice (Vidal et al., 1997). As suggested by Baragiola et al. (1999), condensed oxygen could be localized in patchy regions of Ganymede's surface, which are not representative of the overall surrounding terrain. Ground-based observations pointed out the non-uniform distribution of the condensed $O_2$ in Ganymede's surface (Calvin and Spencer, 1997). Spatially resolved observations with Hubble Space Telescope were used to obtained the first geographic distribution of condensed $O_2$ on the Ganymede's surface (Trumbo et al., 2021). The authors showed that $O_2$ is more concentrated in the trailing hemisphere of the satellite at low-to-mid latitudes, in accord with Spencer et al. (1995) and Calvin and Spencer (1997). Vidal et al. (1997) performed a set of experiments which demonstrated that solid $O_2$ mixed with the more refractory water ice is not retained in sufficient quantity to explain observations of Ganymede's surface during daytime. Therefore, Vidal et al. (1997) suggested that solid $O_2$ may exist in a cold (<50 K) subsurface layer or in an atmospheric haze of Ganymede. Unfortunately, $O_2$ is not stable at the surface of the Galilean satellites neither in the solid form nor as a clathrate (Calvin et al., 1996; Hand et al., 2006) due to the satellites' temperature and pressure conditions. However, a clathrate mixed form with $CO_2$ and $SO_2$ can be much more stable near the ice surface than the pure clathrate and solid $O_2$ (Hand et al., 2006). This would allow the presence of two $O_2$ molecules in the clathrate cage at the surface of the moon contributing to the observed $O_2$ absorption bands in the visible range. Our work suggests that mixtures containing molecular oxygen as well as water and carbon dioxide can reproduce observational reflectance data of the Ganymede's surface better than pure $O_2$ ice in the temperature range 20-35 K. Nevertheless, some open questions on the nature and properties of molecular oxygen at the surface of the Jovian's satellites still remain.

An observational challenge for the upcoming JUICE mission and other Earth-based and space-borne observations will be to investigate the presence of solid $O_2$ on dust and/or haze in the Ganymede's atmosphere. Considering the Ganymede's magnetosphere and its influence in modifying the satellite's surface composition, future observations with JUICE would be highly useful to understand the role of plasma, connection with Io and the open-close field line boundary of Ganymede's magnetosphere at the surface. Future astronomical observations, carried out at high spatial resolution (e.g., a few km per pixels), will be able to evidence a correlation between the local flux of energetic ions and the intensity of the observed (0,0) and (1,0) transition bands of molecular oxygen. Any ice surface variability on spatial scale of the order of few kilometers is well within the MAJIS expected performances, with maximum spatial resolution values comparable or better than this spatial scale for observations taken during most JUICE flybys of Ganymede, and particularly during the Ganymede orbit phase. Similarly, the JANUS (Jovis, Amorum ac Natorum Undique Scrutator) camera (Della Corte et al., 2019), which is part of the JUICE payload, is equipped with 13 filters including a panchromatic one, and will not be able to reveal the two features indicative of the presence of $O_2$. However, the near-UV region is not covered by MAJIS, and only marginally covered by JANUS through a broad-band filter. Hence, any ozone feature in this spectral domain will not be investigated. Solid ozone, a product of the chemical alteration of $O_2$, presents some absorption features in the VNIR wavelength range, such as a transition at about 600 nm and one at 4.75 µm in the IR, that could be studied with the MAJIS spectrometer. New insight on the $O_3$ distribution could be obtained also with the upcoming James Webb Space Telescope (JWST) measurements in the mid-infrared (MIR; Norwood et al., 2016). In this spectral region, the asymmetric stretch of $O_3$ is observed at about 9.66 µm. The observation of ozone ice would imply the presence of solid $O_2$. Detailed comparisons with the ozone distribution, as observed in the MIR range by JWST and in the VNIR by JUICE, could reveal any correlation with oxygen on the surface of the Galilean satellites. The same maps could be compared with previous measurements of $O_3$, obtained with the Galileo spacecraft and HST, to better address the suggested anti-correlation between oxygen and ozone with respect to latitude (Trumbo et al., 2021). The work presented here highlights the importance of combining mid-to-high spectral and high spatial resolution data of the surface of the Jovian satellites with a systematic set of laboratory data with appropriate spectral resolution to guide a correct interpretation of present and future observations.

**Acknowledgements**
The authors thank the anonymous reviewers for their many insightful comments and suggestions. The research presented in this work has been supported by the Royal Society University Research Fellowship (UF130409), the Royal Society Research Fellow Enhancement Award (RGF/EA/180306), and the project CALIPSOplus under the Grant Agreement 730872 from the EU Framework Programme for Research and Innovation HORIZON 2020. The research of Z.K. was supported by VEGA – the Slovak Grant Agency for Science (grant No. 2/0059/22), and COST Action TD 1308 and the Slovak Research and Development Agency (contract No. APVV-19-0072). G.S. was supported by the Italian Space Agency (ASI 2013-056 JUICE Partecipazione Italiana alla fase A/B1) and by the European COST Action TD1308-Origins and evolution of life on Earth and in the Universe (ORIGINS). S.I. recognizes the Royal Society for financial support. A.M. research was partly supported by ASI (scientific contract nr. 2018-25-HH.0).

**Appendix A**

In Figure A1, we compare the observed spectrum of Ganymede obtained at TNG with some of the laboratory data selected according to the results shown in Figs. 5 and 6 to determine whether the observed band profiles can be reproduced by a set of laboratory data. Figure A1 is analog to Fig. 7 in the main text and presents the residual spectra for all the selected comparisons. All selected spectra of pure and mixed $O_2$-rich ice can reproduce the observed (1,0) band profile at 576.1 nm. However, the observed (0,0) band at 624.0 nm is noisier due to sub-optimal removal of telluric signal, and seems to be broader than any laboratory spectra presented here. Results from Fig. A1 are in good agreement with those shown in Fig. 7 that compares the same laboratory spectra to observational data from Spencer et al. (1995). We confirm that selected mixtures containing $H_2O$, $N_2$, and $CO_2$ can reproduce observational data more accurately (minimal residual spectra) than a spectrum of pure $O_2$, regardless of its temperature.

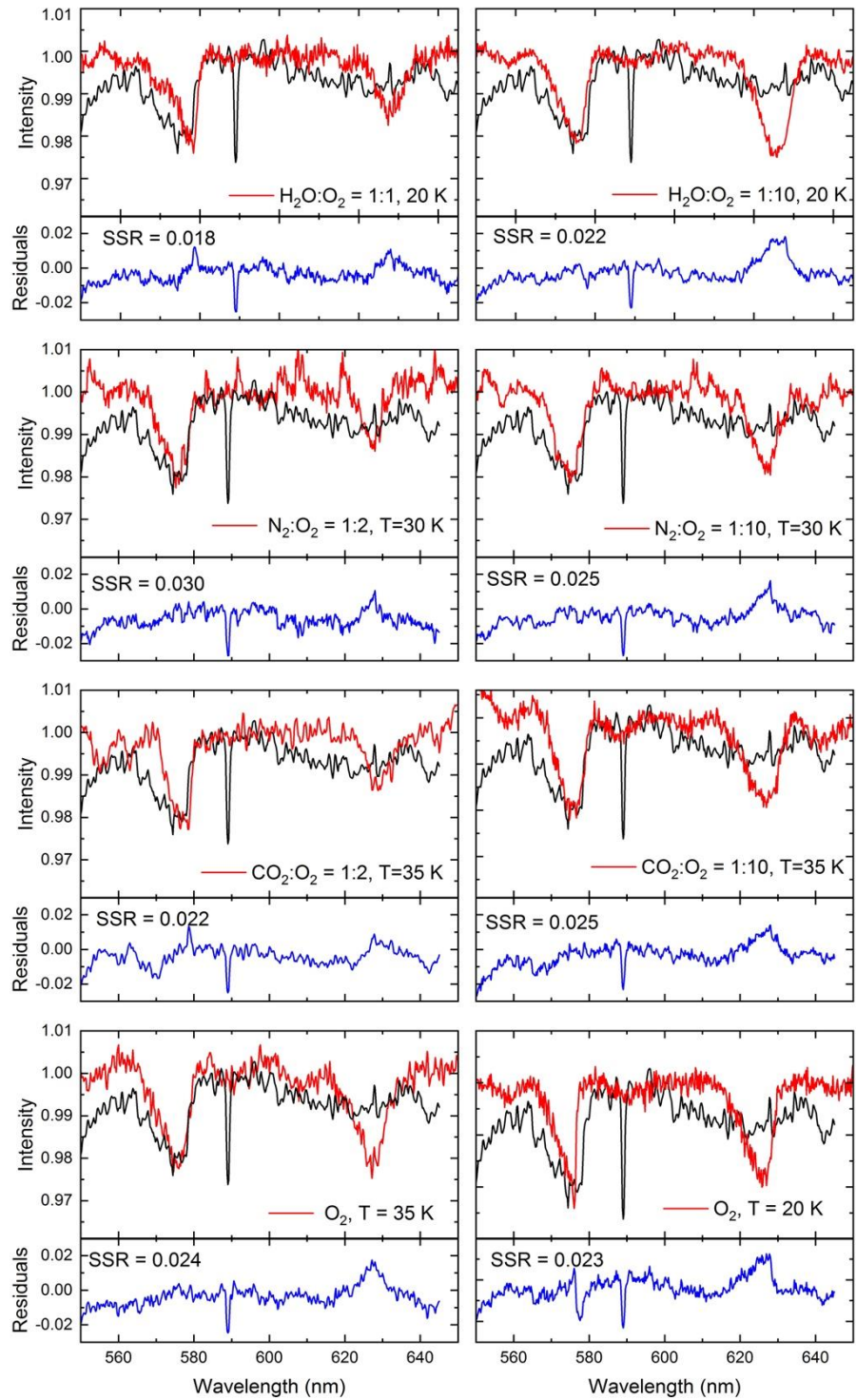

**Fig. A1.** Ground-based spectrum of Ganymede acquired in reflectance at TNG (black curve) compared to selected laboratory transmittance spectra of solid $O_2$, pure and mixed with $H_2O$, $CO_2$, and $N_2$, with different mixing ratios and ice temperatures (in red). Blue lines are the spectral residue that is the difference between black and red lines. SSR stands for sum of squared residuals.